\documentclass[iop]{emulateapj}
\usepackage{nicefrac}
\usepackage{amsmath}
\usepackage[colorlinks=true, urlcolor=magenta, citecolor=blue]{hyperref}
\usepackage{graphicx}
\usepackage{mathtools}
\usepackage{enumitem}
\usepackage{gensymb}

\newcommand{\myemail}{\href{mailto:arcavi@ucsb.edu}{arcavi@ucsb.edu}}

\def\gkg{SN\,2016gkg}
\def\Msun{{\rm M}_\odot}

\def\Rsun{{\rm R}_\odot}

\newcommand{\ra}[4]{$#1^{\rm h}#2^{\rm m}#3^{\rm s}\mathllap{.}#4$}
\newcommand{\dec}[4]{$#1\degree#2'#3'\mathclap{.}'#4$}

\newcounter{Affiliation}
\newcommand{\aftext}[1]{\refstepcounter{Affiliation}\altaffilmark{\theAffiliation}#1}

\shorttitle{{\gkg} Shock Cooling}
\shortauthors{Arcavi et al.}

\begin{document} 


\title{Constraints on the Progenitor of SN\,2016gkg From Its Shock-Cooling Light Curve}


\author{Iair~Arcavi\altaffilmark{\ref{af:UCSB},\ref{af:LCOGT},$\star$},
Griffin~Hosseinzadeh\altaffilmark{\ref{af:LCOGT},\ref{af:UCSB}},
Peter~J.~Brown\altaffilmark{\ref{af:TexasAM}},
Stephen~J.~Smartt\altaffilmark{\ref{af:QUB}},
Stefano~Valenti\altaffilmark{\ref{af:Davis}},
Leonardo~Tartaglia\altaffilmark{\ref{af:TexasTech},\ref{af:Davis}},
Anthony~L.~Piro\altaffilmark{\ref{af:Carnegie}},
Jos\'{e}~L.~Sanchez\altaffilmark{\ref{af:Geminis}},
Brent~Nicholls\altaffilmark{\ref{af:MtVernon}},
Berto~L.A.G.~Monard\altaffilmark{\ref{af:Kleinkaroo}},
D.~Andrew~Howell\altaffilmark{\ref{af:LCOGT},\ref{af:UCSB}},
Curtis~McCully\altaffilmark{\ref{af:LCOGT},\ref{af:UCSB}},
David~J.~Sand\altaffilmark{\ref{af:TexasTech}},
John~Tonry\altaffilmark{\ref{af:IfA}}, 
Larry~Denneau\altaffilmark{\ref{af:IfA}}, 
Brian~Stalder\altaffilmark{\ref{af:IfA}}, 
Ari~Heinze\altaffilmark{\ref{af:IfA}}, 
Armin~Rest\altaffilmark{\ref{af:STScI}}, 
Ken~W.~Smith\altaffilmark{\ref{af:QUB}}
and
David~Bishop\altaffilmark{\ref{af:Rochester}}
}

\affil{
\aftext{Department of Physics, University of California, Santa Barbara, CA 93106-9530, USA\label{af:UCSB}; \myemail}\\
\aftext{Las Cumbres Observatory Global Telescope, 6740 Cortona Dr Ste 102, Goleta, CA 93117-5575, USA\label{af:LCOGT}}\\
\aftext{George P. and Cynthia Woods Mitchell Institute for Fundamental Physics \& Astronomy, Department of Physics and Astronomy, Texas A \& M University, College Station, USA\label{af:TexasAM}}\\
\aftext{Astrophysics Research Centre, School of Mathematics and Physics, Queen’s University Belfast, Belfast BT7 1NN, UK\label{af:QUB}}\\
\aftext{Department of Physics, University of California, 1 Shields Ave, Davis, CA 95616-5270, USA\label{af:Davis}}\\
\aftext{Physics \& Astronomy Department, Texas Tech University, Lubbock, TX 79409, USA\label{af:TexasTech}}\\
\aftext{Carnegie Observatories, 813 Santa Barbara Street, Pasadena, CA 91101, USA\label{af:Carnegie}}\\
\aftext{Observatorio Astronomico Geminis Austral, Rosario, Argentina\label{af:Geminis}}\\
\aftext{Mt. Vernon Observatory, 6 Mt. Vernon pl, Nelson, New Zealand\label{af:MtVernon}}\\
\aftext{Kleinkaroo Observatory, Calitzdorp, Western Cape, South Africa\label{af:Kleinkaroo}}\\
\aftext{Institute for Astronomy, University of Hawaii, 2680 Woodlawn Drive Honolulu, HI 96822, USA\label{af:IfA}}\\
\aftext{Space Telescope Science Institute, 3700 San Martin Drive, Baltimore, MD 21218, USA\label{af:STScI}}\\
\aftext{Rochester Academy of Science, P.O. Box 92642, Rochester, New York 14692-0642, USA\label{af:Rochester}}\\
}
\altaffiltext{$\star$}{Einstein Fellow}



\newpage

\begin{abstract} 

{\gkg} is a nearby Type~IIb supernova discovered shortly after explosion. Like several other Type~IIb events with early-time data, {\gkg} displays a double-peaked light curve, with the first peak associated with the cooling of a low-mass extended progenitor envelope. We present unprecedented intranight-cadence multi-band photometric coverage of the first light-curve peak of {\gkg} obtained from the Las Cumbres Observatory Global Telescope network, the Asteroid Terrestrial-impact Last Alert System, the {\it Swift} satellite and various amateur-operated telescopes. Fitting these data to analytical shock-cooling models gives a progenitor radius of ${\sim}40$--$150\,\Rsun$ with ${\sim}2$--$40\times10^{-2}\,\Msun$ of material in the extended envelope (depending on the model and the assumed host-galaxy extinction). Our radius estimates are broadly consistent with values derived independently (in other works) from {\it HST} imaging of the progenitor star. However, the shock-cooling model radii are on the lower end of the values indicated by pre-explosion imaging. Hydrodynamical simulations could refine the progenitor parameters deduced from the shock-cooling emission and test the analytical models.

\smallskip
\end{abstract} 




\section{Introduction} 
\setcounter{footnote}{0}

Type~IIb supernovae (SNe) are a class of explosions defined by the presence of hydrogen in their spectra (similar to Type~II SNe) at early times and strong helium lines (similar to Type~Ib SNe) at later times. Such spectral evolution suggests that Type~IIb SN progenitors are partially stripped stars, having lost some but not all of their hydrogen envelope. 

Curiously, some Type~IIb SNe show double-peaked light curves \citep{Richmond1994,Arcavi2011,Kumar2013,Bufano2014,Morales-Garoffolo2014}, apparent in all optical bands. The common interpretation is that the first peak is due to the cooling of the ejecta following the shock breakout, while the second peak is from nickel-decay power.

The first double-peaked Type~IIb SN to have been studied with modern shock-cooling models was SN\,2011dh \citep{Arcavi2011}. Its rapid temperature evolution, when compared to the \citet{Rabinak2011} models, suggested a compact ($R{\sim}10^{11}$\,cm) progenitor, while pre-explosion {\it HST} images indicated an extended ($R{\sim}10^{13}$\,cm) progenitor \citep{Maund2011, VanDyk2011, VanDyk2013}. This discrepancy was later attributed to the fact that the \citet{Rabinak2011} models consider red supergiants (RSGs) and blue supergiants (BSGs) with massive envelopes, while the progenitors of double-peaked IIb SNe likely have a different structure. 

\citet{Bersten2012} showed that hydrodynamical modeling of SN\,2011dh as the explosion of a star with a compact core and a low-mass extended envelope can reproduce the full double-peaked light curve, its fast temperature evolution and a progenitor radius consistent with that inferred from pre-explosion {\it HST} imaging (such a structure had in-fact already been suggested for the progenitor star of the first double-peaked Type~IIb SN\,1993J by \citealp{Hoflich1993} and \citealp{Woosley1994a}).

\citet[][hereafter NP14]{Nakar2014} later confirmed that progenitors with massive envelopes (which they call ``standard progenitors'') can not produce double-peaked light curves in the redder (e.g $R$ and $I$) bands and that stars with low-mass extended envelopes (which they call ``non-standard progenitors'') are indeed required. NP14 also reproduced the \citet{Bersten2012} progenitor parameters for SN\,2011dh using approximate analytical expressions linking the time and luminosity of the first peak to the pre-explosion mass and radius of the extended material. \citet[][hereafter P15]{Piro2015} later expanded the NP14 model, providing an analytical expression for the full light curve shape around the first peak. Both NP14 and P15 assume that the extended mass is lower than the core mass and that it is concentrated around the outer radius of the envelope, but make no explicit assumptions on the precise density profile, whether it is polytropic or not. More recently, \citet[][hereafter SW16]{Sapir2016} extended the analytical model of \citet{Rabinak2011}, which specifically considers polytropic density profiles, to later times by calibrating it to numerical simulations. Unlike the original \citet{Rabinak2011} model, the SW16 extension can produce double-peaked light curves in all optical bands, for low enough envelope masses. 

Here we present early-time multi-band observations of the Type~IIb {\gkg}, covering its shock-cooling peak in unprecedented detail. {\gkg} was discovered by A.~Buso at $\alpha\textrm{(J2000)}$=\ra{01}{34}{14}{46} and $\delta\textrm{(J2000)}$=\dec{-29}{26}{25}{0} on Sep 20.19 UT\footnote{\url{http://ooruri.kusastro.kyoto-u.ac.jp/mailarchive/vsnet-alert/20188}} and reported by A.~Buso and S.~Otero\footnote{\url{https://wis-tns.weizmann.ac.il/object/2016gkg}}. Shortly after discovery, the transient was confirmed by the All Sky Automatic Survey for SNe \citep[ASAS-SN;][]{Nicholls2016} and by the Asteroid Terrestrial-impact Last Alert System \citep[ATLAS;][]{Tonry2011, Tonry2016}. {\gkg} was initially classified as a Type~II SN owing to broad H$\alpha$ and H$\beta$ P-Cygni features in its spectrum \citep{Jha2016}. The emergence of broad He~I features later refined the classification to that of a Type~IIb SN \citep{VanDyk2016}.

Pre-explosion imaging of the SN site from {\it HST} \citep{Kilpatrick2016a} has been used to derive a progenitor radius estimate of $138_{-103}^{+131}\,\Rsun$ \citep{Kilpatrick2016b} and  $\sim150$--$320\,\Rsun$ \citep[][considering both their progenitor candidates]{Tartaglia2016}. \citet{Kilpatrick2016b} further find that a broad progenitor radius range of $257_{-189}^{+389}\,\Rsun$ is consistent with the rise to the first light curve peak using the \citet{Rabinak2011} model. \citet{Tartaglia2016} fit this model to the first few days of the temperature evolution (rather than the luminosity evolution) and find it to be consistent with a progenitor radius of $\sim48-124\,\Rsun$. 

In this work we fit the NP14, P15 and the SW16 models, which are better suited for double-peaked SNe IIb, to the early-time light curve of {\gkg} in order to test these models and obtain an independent estimate of the progenitor and explosion properties.

\section{Observations}

\begin{figure*}
\includegraphics[width=\textwidth]{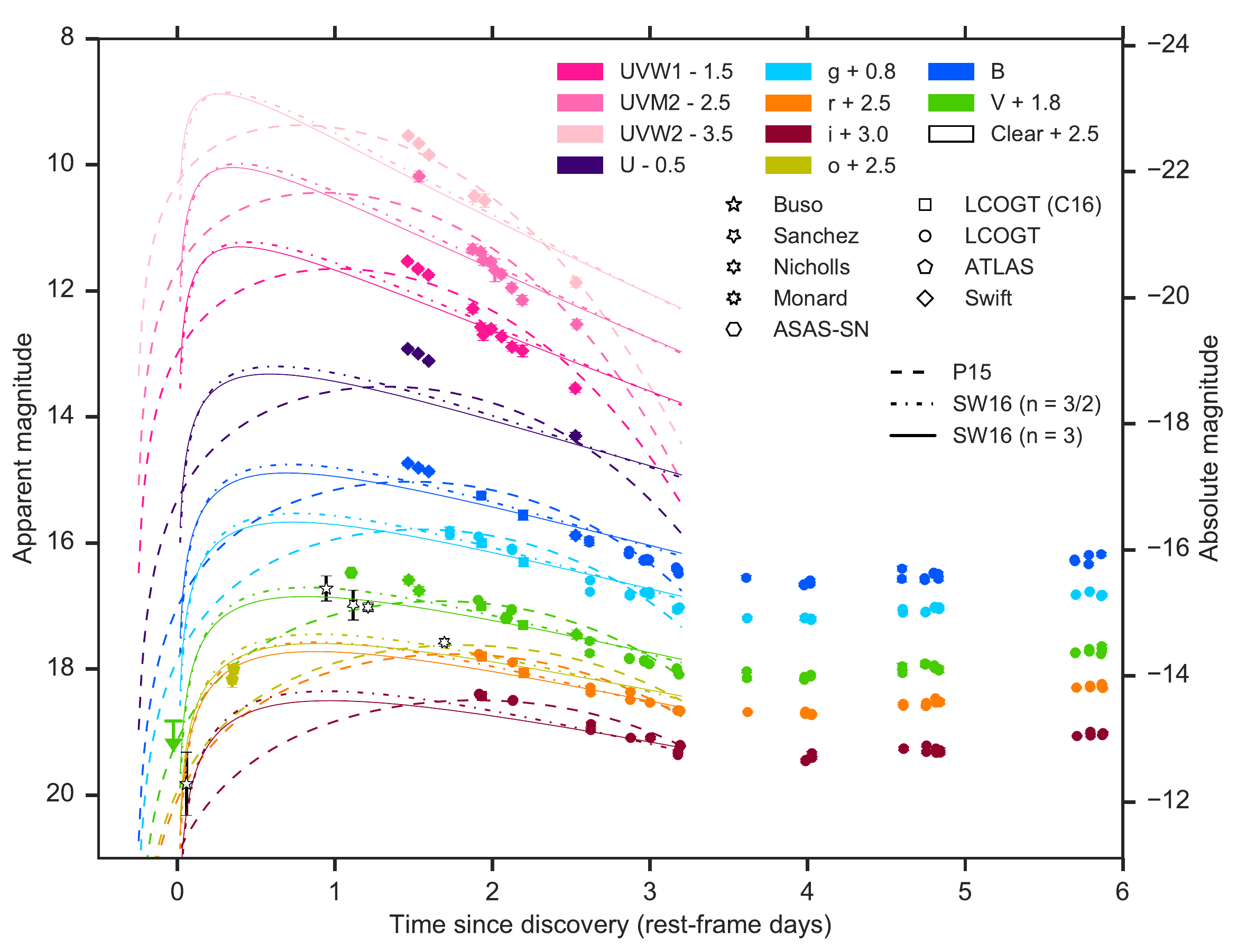}
\caption{\label{fig:lc}Early-time photometry of {\gkg} corrected for Milky-Way extinction and a host-galaxy extinction value of E(B-V)=0.09 \citep{Tartaglia2016}. Best-fit P15 and SW16 models to the first peak are shown in lines. The Buso \& Otero (stars), ASAS-SN (hexagons and upper limit arrow) and publicly available LCOGT data (squares) are taken from \citet{Chen2016}. Error bars denote $1\sigma$ uncertainties (they are sometimes smaller than the data markers). The SW16 models are plotted for their respective validity time ranges (Eq. \ref{eq:sw16tlimits}). (Supplementary data files of this figure are available in the online journal.)}
\end{figure*}

We compile data from the discovery report by A. Buso \& S. Otero, publicly-available followup observations taken with the Las Cumbres Observatory Global Telescope (LCOGT) network and ASAS-SN \citep{Chen2016} and our reduction of publicly-available {\it Swift} UVOT data for the first light-curve peak, to which we add observations taken with 30--40\,cm amateur-operated telescopes, our ATLAS early-time detections and our own intensive followup campaign with LCOGT.

ATLAS is a twin 0.5-m telescope system on Haleakala and Mauna Loa. The first unit is operational on Haleakala and during the course of its robotic survey operations detected {\gkg} in two 30-second exposures 9.05 and 9.42 hours after discovery. The images were taken in the ATLAS orange filter (denoted $o$), which is a broad rectangular bandpass covering $5600$--$8200$\,\AA\ and is primarily used during bright moon time. Image reductions were carried out with a customized pipeline, with zeropoint calibration calculated for each frame based on a custom catalog from Pan-STARRS1 \citep{Magnier2013,Schlafly2013}. A reference sky frame was subtracted from the two target frames using a custom pipeline based on HOTPANTS and point-spread-function (PSF) photometry was measured using Waussian profiles (see \citealt{Tonry2011} for a comparison between profiles). 

Images by J.~Sanchez were taken through a 40.6-cm Skywatcher Newton f/4 telescope with a ZWO ASI1600 MM-C camera fitted with a ZWO UV IR CUT filter, which is a broad rectangular bandpass covering $\sim4000$--$7000$\,\AA)\footnote{\url{https://astronomy-imaging-camera.com/wp-content/uploads/IR-Window-graph1.jpg}}. Exposures were dark and flat corrected and then stacked. Images by B.~Nicholls were taken through a 30-cm Meade LX200 f/10 telescope, which is in trust with C.~Rowe \& D.~Victor, with an unfiltered ATIK 4000le camera. Exposures were stacked after dark subtractions. Images by B.~Monard were taken through a 30-cm Meade RCX400 f/8 telescope with an unfiltered SBIG ST8-XME camera. Individual images were dark subtracted and flat fielded, and selective stacking was applied in median mode. Aperture photometry was used to measure the flux of the SN in the stack-combined image from each telescope. The photometry was then calibrated to $r$-band using stars from the APASS catalog \citep{Henden2009}. 

LCOGT is a worldwide network of robotic 0.4-m, 1-m and 2-m telescopes designed for high-cadence rapid response to transient events \citep{Brown2013a}. Images of {\gkg} were obtained using the LCOGT 1-m telescopes at the South African Astronomical Observatory, the Cerro-Tololo Inter-American Observatory in Chile and Siding Spring observatory in Australia using both SBIG and Sinistro cameras. Initial processing of the images was performed using the custom Python-based {\tt BANZAI} pipeline. Photometry was then extracted using the PyRAF-based {\tt lcogtsnpipe} pipeline \citep{Valenti2016} to perform PSF fitting and photometry extraction. The $BVgri$-band photometry was calibrated to the APASS catalog. The $U$-band photometry was calibrated to standard fields \citep{Stetson2000} observed on the same night as the SN field.

The {\it Swift} UVOT \citep{Gehrels2004a,Roming2005} data were reduced using the pipeline of the {\it Swift} Optical/Ultraviolet Supernova Archive \citep[SOUSA;][]{Brown2014}. The reduction is based on that of \citet{Brown2009} using the revised UV zeropoints and time-dependent sensitivity from \citet{Breeveld2011}.

We adopt a distance of 26.4\,Mpc and a distance modulus of 32.11 magnitudes to {\gkg}, based on Tully-Fisher distance measurements \citep{Tully2009} to its host galaxy, NGC 613, retrieved via the NASA Extragalactic Database (NED). We correct for Milky Way extinction using the dust maps of \citet{Schlafly2011}, obtained via NED, for the $UBVgri$ filters (we use the $R$-band extinction values for correcting the Clear and $o$-band data). We use the \citet{Cardelli1989} law with $R_V=3.1$ to correct the {\it Swift} $UVW1$, $UVM2$ and $UVW2$ magnitudes. From the Na I D EW values measured by \citet{Tartaglia2016} in a high-resolution spectrum, and using the conversions of \citet{Poznanski2012}, we adopt a host-galaxy extinction of ${\rm E(B-V)}=0.09_{-0.03}^{+0.06}$.

Our data is presented in Figure \ref{fig:lc}, where ground-based $UBV$ magnitudes are in the Vega system and $gori$ magnitudes are in the AB system. The {\it Swift} data are presented in the UVOT system \citep{Poole2008,Breeveld2011} and can be found also on SOUSA.

\section{Analysis}

\begin{figure*}
\includegraphics[width=0.33\textwidth]{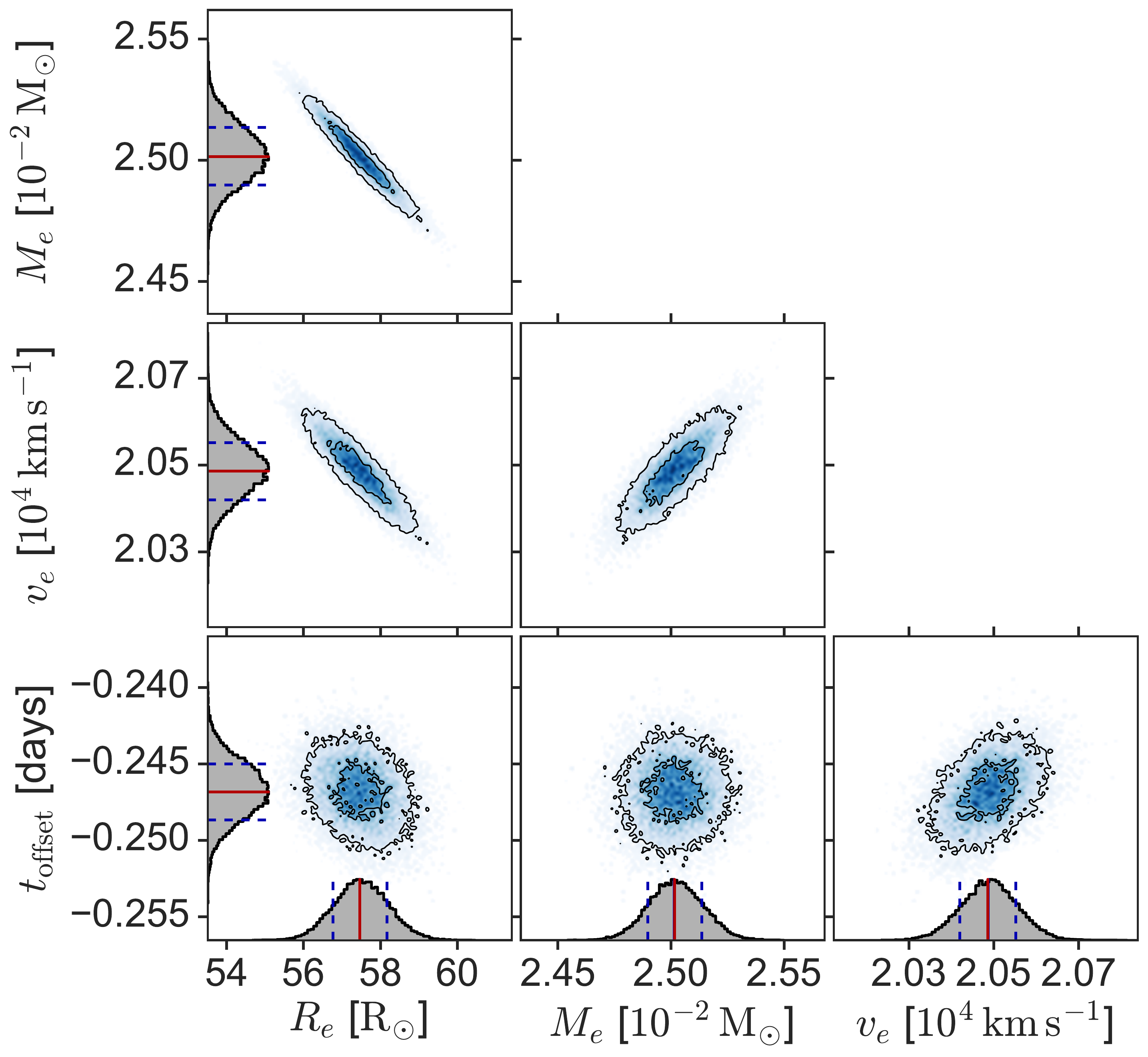}\,\,\includegraphics[width=0.33\textwidth]{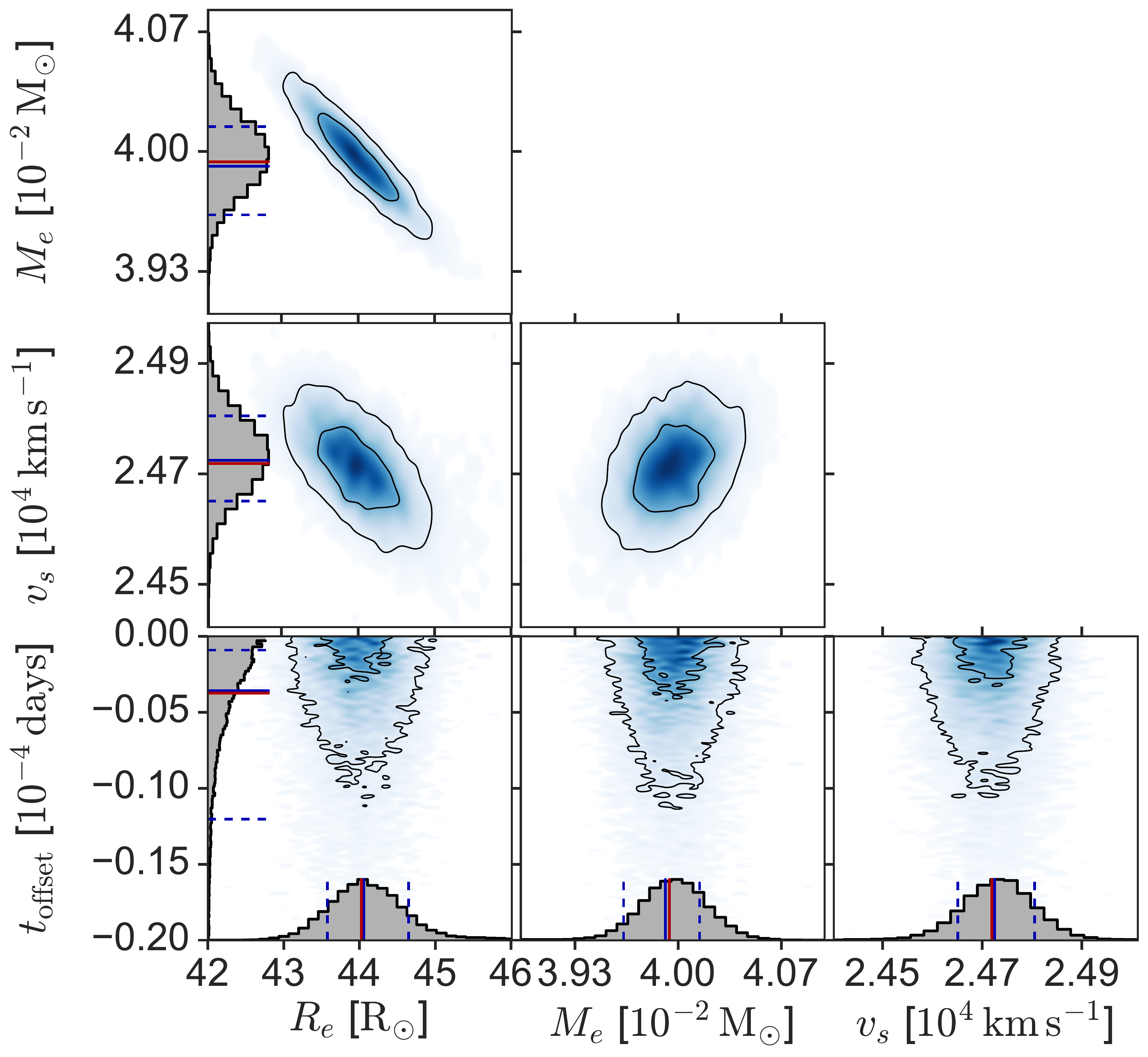}\,\,\includegraphics[width=0.33\textwidth]{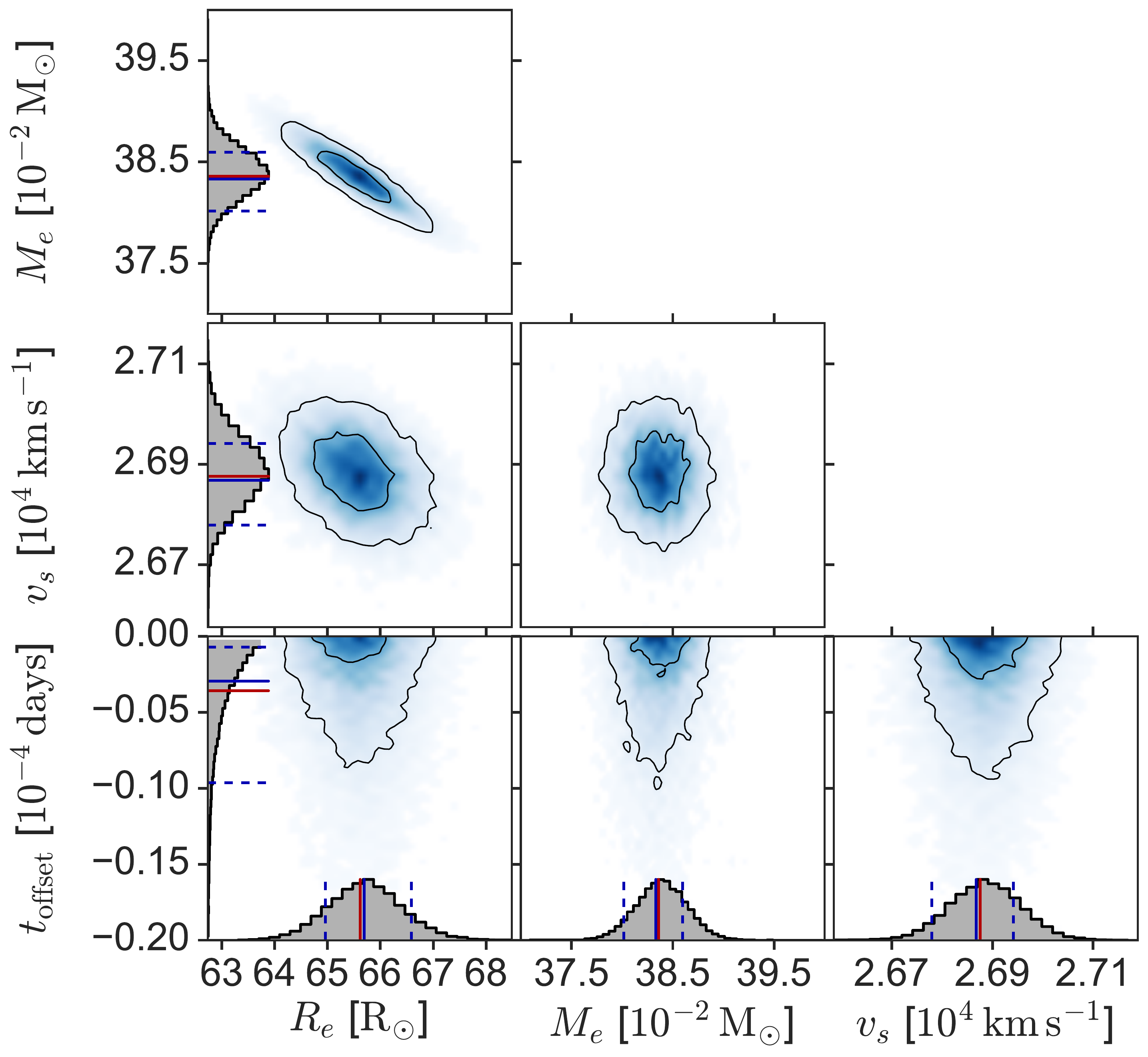}
\caption{\label{fig:corner}Parameter distribution for the model fits (left to right: P15, SW16 [$n=1.5$], SW16 [$n=3$]) assuming the nominal host extinction value of E(B-V)=0.09. The contour lines denote 50\% and 90\% bounds. The red and blue solid lines over-plotted on each histogram denote the mean and median of each parameter distribution (respectively). The dashed lines denote 68\% confidence bounds.}
\end{figure*}

We consider three analytical models of the shock cooling emission: NP14, which connects the time and luminosity of the first light curve peak to the radius and mass of the extended envelope; P15, which extends this model to an expression of the early-time light curve peak; and SW16 which extends the \citet{Rabinak2011} models to an expression of the early-time light curve peak.

\subsection{NP14 Model}

NP14 provide a relation between the luminosity of the first peak and the radius $R_e$ of the extended envelope of the progenitor star, and between the time of the first peak and the mass $M_e$ concentrated around $R_e$:
\begin{equation} \label{eq:r_ext}
R_{e} \approx 2\times10^{13}\kappa_{0.34}\left[\frac{L_{bol}\left(t_p\right)}{10^{43}\textrm{\,erg\,s}^{-1}}\right]v_9^{-2}\textrm{\,cm}
\end{equation}
\begin{equation} \label{eq:m_ext}
M_{e} \approx 5\times10^{-3}\kappa_{0.34}^{-1}v_{9}\left(\frac{t_p}{1\textrm{\,day}}\right)^{2}\textrm{\,}\Msun
\end{equation}
where $\kappa_{0.34}$ is the opacity in units of $0.34$\,cm$^2$\,g$^{-1}$, $v_{9}$ is the characteristic velocity $v_e$ of the extended material in units of $10^9\textrm{\,cm\,s}^{-1}$ and $L_{bol}\left(t_p\right)$ is the bolometric luminosity at the time of first peak $t_p$. 

We fit the A.~Buso and $V$-band data with a smoothing spline to interpolate the time and luminosity of the first peak. We set the explosion time to that found by the best P15 fit (see below). Since we do not have multi-band coverage at $t_p$, we approximate $L_{bol}\left(t_p\right)$ as the interpolated $V$-band peak luminosity times a bolometric correction, assuming a blackbody with temperature $T_{obs}$ at peak. An estimate of this temperature, as given by NP14, is:
\begin{equation} \label{eq:t_bb}
T_{obs}\left(t_p\right) \approx 3\times10^{4}\kappa_{0.34}^{-1/4}\left(\frac{t_p}{1\textrm{\,day}}\right)^{-1/2}R_{13}^{1/4}\textrm{\,K}
\end{equation}
where $R_{13} = R_{e}/10^{13}$\,cm. 

Since $L_{bol}\left(t_p\right)$ is required in order to obtain $R_{e}$, $T_{obs}\left(t_p\right)$ is required in order to obtain $L_{bol}\left(t_p\right)$, and $R_{e}$ is required to obtain $T_{obs}\left(t_p\right)$, we solve for $R_{e}$, $L_{bol}\left(t_p\right)$ and $T_{obs}\left(t_p\right)$ iteratively until the difference in the values of $R_{e}$ and $T_{obs}\left(t_p\right)$ between consecutive iterations is less than $1\%$. The solution converges within a few iterations and is not sensitive to the initial values used.

This model assumes one characteristic velocity for the extended material, $v_e$. In reality, there is a velocity gradient in the envelope. \citet{Jha2016} measure an expansion velocity of $1.7\times10^9$\,cm\,s$^{-1}$ from the minimum of the H$\alpha$ P-cygni line in the first {\gkg} spectrum obtained 1.7 days from discovery. We set $v_e$ to this value, though the exact relation between $v_e$ and the measured expansion velocity of H depends on the velocity profile of the expanding envelope, which is not known.

\subsection{P15 Model}

We cast the P15 analytical expression for the shape of the first light curve peak (their Eq. 15) in terms of $M_{e}$, $R_{e}$, $v_9$ and the mass of the core $M_c$:
\begin{align} \label{eq:p15lum}
L\left(t\right)\approx& 8.27\times10^{42}{\kappa}_{0.34}^{-1}v_9^2R_{13}\left(\frac{M_c}{\Msun}\right)^{0.01}\times \nonumber \\
& \exp\left[-4.135\times10^{-11}t\left(tv_9+2\times10^{4}R_{13}\right) \vphantom{ \frac{M_{e}}{0.01\Msun}^{-1}} \times \right. \nonumber \\
& \left. {\kappa}_{0.34}^{-1}\left(\frac{M_c}{\Msun}\right)^{0.01}\left(\frac{M_{e}}{0.01\Msun}\right)^{-1}\right]\textrm{\,erg\,s}^{-1}
\end{align}
where $t$ is the time since explosion in seconds. We set $M_c=\Msun$ (the dependence on this parameter is very weak). 

Following P15 we assume the emission is a blackbody at radius $R\left(t\right)=R_{e}+v_{e}t$. This allows us to estimate the temperature:
\begin{equation} \label{eq:p15temperature}
T\left(t\right)=\left[\frac{L\left(t\right)}{4{\pi}R^2\left(t\right){\sigma}_{SB}}\right]^{1/4}
\end{equation}
(with ${\sigma}_{SB}$ the Stefan-Boltzmann constant) and thus the luminosity in any band, given $R_{e}$, $M_{e}$, $v_{e}$ and the explosion time. 

\subsection{SW16 Model}

SW16 extend the analytical models of \citet{Rabinak2011} for the temperature and luminosity of the shock-cooling emission out to a few days post-explosion. At these phases, as the radiation emerges from inner layers, the self-similar solution determined by \citet{Rabinak2011} is no longer valid. Instead, for low enough envelope masses, SW16 find that the luminosity is suppressed, producing an early-time light curve peak. 

The final expression for the luminosity according to SW16 is:
\begin{align} \label{eq:sw16lum}
L\left(t\right) = & 1.88\left[1.66\right]\times10^{42} \times \nonumber \\ 
& \left(\frac{v_{s,8.5}t^2}{f_{\rho}M\kappa_{0.34}}\right)^{-0.086\left[-0.175\right]}\frac{v_{s,8.5}^2R_{13}}{\kappa_{0.34}} \times \nonumber \\
& \exp\left\{ -\left[\frac{1.67\left[4.57\right]t}{\left(19.5\kappa_{0.34}M_ev_{s,8.5}^{-1}\right)^{0.5}}\right]^{0.8\left[0.73\right]} \right\} \textrm{\,erg\,s}^{-1}
\end{align}
for a polytropic index of $n=3/2\left[3\right]$ typical of stars with convective envelopes such as RSGs [stars with radiative envelopes such as BSGs], where $t$ here is in days, $M_e$ and $M=M_e+M_c$ are in solar masses, $v_{s,8.5}$ is the velocity of the shock $v_s$ in units of $10^{8.5}$\,cm\,s$^{-1}$ and 
\begin{equation} \label{eq:sw16f}
f_{\rho}\approx\begin{cases}
\left(M_{e}/M_{c}\right)^{0.5} & n=3/2\\
0.08\left(M_{e}/M_{c}\right) & n=3
\end{cases}
\end{equation}
As in the P15 fits, we fix $M_c=\Msun$. 

The color temperature is given by:
\begin{align} \label{eq:sw16temperature}
T\left(t\right) \approx & 2.05\left[1.96\right] \times10^4 \times \nonumber \\
& \left(\frac{v_{s,8.5}^2t^2}{f_{\rho}M\kappa_{0.34}}\right)^{0.027\left[0.016\right]} \frac{R_{13}^{0.25}}{\kappa_{0.34}^{0.25}}t^{-0.5} \textrm{K}
\end{align}
where $t$ is in days and $M$ is in solar masses. This, again, allows us to fit any band as a function of $R_{e}$, $M_{e}$, $v_{s}$ and the explosion time. 

According to SW16, the validity of this model is limited to times:
\begin{align} \label{eq:sw16tlimits}
t > & 0.2 \frac{R_{13}}{v_{s,8.5}}\max\left[0.5,\frac{R_{13}^{0.4}}{\left(f_{\rho}\kappa_{0.34}M\right)^{0.2}v_{s,8.5}^{0.7}}\right]\,\textrm{days} \\
t < & 7.4 \left(\frac{R_{13}}{\kappa_{0.34}}\right)^{0.55}\,\textrm{days}
\end{align} 

\renewcommand{\arraystretch}{1.8}
\begin{table*}
{\caption{\label{tab:results}Best fit parameters to the early-time light curve peak from the NP14, P15 and SW16 models, assuming the nominal, lower and upper values for the host extinction. Errors denote $68\%$ confidence bounds. The SW16 models prefer an explosion time consistent with the discovery epoch.}}
\center
\begin{tabular}{lccccc}
\hline
\hline
Parameter & E(B-V) & NP14 & P15 & \multicolumn{2}{c}{SW16} \\
 & & & & ($n=3/2$) & ($n=3$) \\
\hline
$R_e$ [$10^{12}$\,cm] & 0.06 & $8.64_{-0.34}^{+0.34}$ & $2.90_{-0.03}^{+0.03}$ & $2.47_{-0.02}^{+0.02}$ & $3.70_{-0.04}^{+0.04}$ \\
 & 0.09 & $9.13_{-0.34}^{+0.34}$ & $4.00_{-0.05}^{+0.05}$ & $3.07_{-0.03}^{+0.04}$ & $4.57_{-0.05}^{+0.06}$ \\
 & 0.15 & $10.15_{-0.13}^{+0.13}$ & $8.03_{-0.11}^{+0.11}$ & $4.90_{-0.06}^{+0.06}$ & $7.30_{-0.10}^{+0.52}$ \\
\hline
$R_e$ [$\Rsun$] & 0.06 & $124.3_{-4.9}^{+4.9}$ & $41.8_{-0.4}^{+0.5}$ & $35.6_{-0.3}^{+0.3}$ & $53.3_{-0.5}^{+0.6}$ \\
 & 0.09 & $131.3_{-4.9}^{+4.9}$ & $57.5_{-0.7}^{+0.7}$ & $44.1_{-0.5}^{+0.6}$ & $65.7_{-0.7}^{+0.9}$ \\
 & 0.15 & $146.0_{-1.8}^{+1.8}$ & $115.5_{-1.5}^{+1.6}$ & $70.5_{-0.8}^{+0.8}$ & $105.0_{-1.4}^{+7.5}$ \\
\hline
$M_e$ [$10^{-2}\,\Msun$] & 0.06 & $1.60_{-0.05}^{+0.05}$ & $2.72_{-0.01}^{+0.01}$ & $4.22_{-0.02}^{+0.02}$ & $40.96_{-0.29}^{+0.27}$ \\
 & 0.09 & $1.60_{-0.05}^{+0.05}$ & $2.50_{-0.01}^{+0.01}$ & $3.99_{-0.03}^{+0.02}$ & $38.34_{-0.30}^{+0.26}$ \\
 & 0.15 & $1.62_{-0.05}^{+0.05}$ & $2.11_{-0.01}^{+0.01}$ & $3.58_{-0.02}^{+0.02}$ & $33.58_{-1.14}^{+0.23}$ \\
\hline
$t_{\rm offset}$ [days] & 0.06 & n/a & $-0.245_{-0.002}^{+0.002}$ & $0$ & $0$ \\
 & 0.09 & n/a & $-0.247_{-0.002}^{+0.002}$ & $0$ & $0$ \\
 & 0.15 & n/a & $-0.236_{-0.002}^{+0.002}$ & $0$ & $0$ \\
\hline
$v_e$ [$10^9\,$cm$\,$s$^{-1}$] & 0.06 & $1.7$ (fixed) & $2.11_{-0.01}^{+0.01}$ & $2.45_{-0.01}^{+0.01}$ & $2.64_{-0.01}^{+0.01}$ \\
 & 0.09 & $1.7$ (fixed) & $2.05_{-0.01}^{+0.01}$ & $2.47_{-0.01}^{+0.01}$ & $2.69_{-0.01}^{+0.01}$ \\
 & 0.15 & $1.7$ (fixed) & $1.92_{-0.01}^{+0.01}$ & $2.49_{-0.01}^{+0.01}$ & $2.76_{-0.01}^{+0.01}$ \\
\hline
\end{tabular}
\end{table*}
\renewcommand{\arraystretch}{1}

For all model fits we set $\kappa_{0.34}=1$, as appropriate for electron scattering of solar composition material. For the P15 and SW16 models we simultaneously fit all bands (fitting the Clear data to the $r$-band model magnitudes) out to 3.2 days from discovery, using the Markov Chain Monte Carlo method through the Python {\tt emcee} package \citep{Foreman-Mackey2012} with $R_{e}$, $M_{e}$, $v_{e}$ (or $v_{s}$ for the SW16 models) and $t_{\rm offset}$ (the offset between discovery and explosion) as free parameters. We allow the explosion time to vary between the discovery date and 0.5 days before the discovery date. We weight the data according to their uncertainties and linearly with time since explosion (with points closer to discovery given larger weights in order to offset the larger number of points post-peak). 

We repeat the fit for three values of host-extinction: the nominal value of E(B-V)=0.09, and the lower and upper bounds of E(B-V)=0.06 and E(B-V)=0.15.

By performing spectrophotometry on the models to compare with the photometric observations, we implicitly take the characteristics of each filter transmission into account \citep{Brown2016}, including the optical tails (known as the ``red leaks'') of the UVOT $UVW2$ and $UVW1$ filters.

\section{Results and Discussion}

\begin{figure}
\includegraphics[width=\columnwidth]{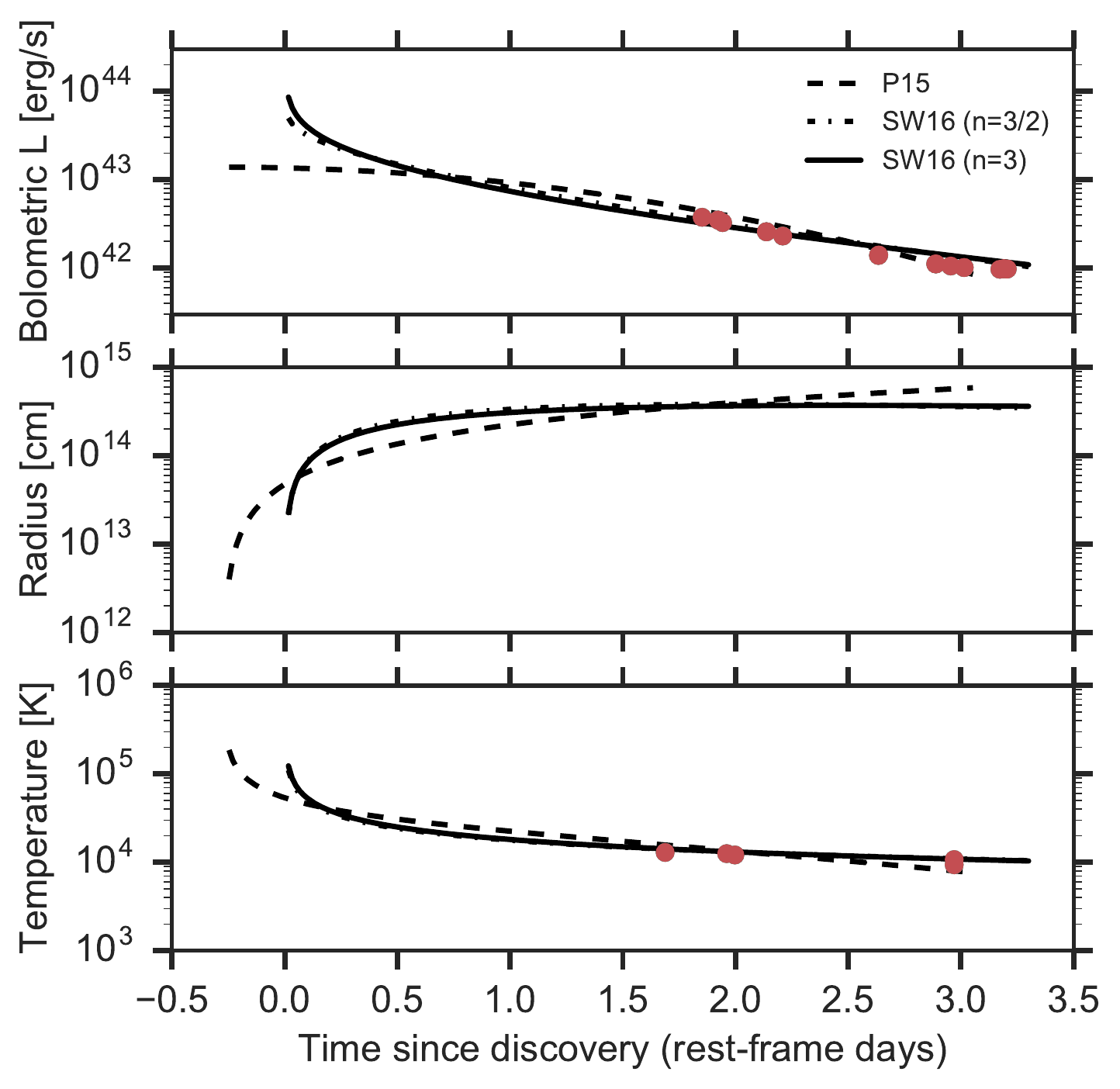}
\caption{\label{fig:lrt}Bolometric luminosity (top), blackbody radius (middle) and temperature (bottom) from the best-fit P15 and SW16 models to the first light-curve peak of {\gkg} assuming the nominal host-galaxy extinction value of E(B-V)=0.09. The points are the pseudo-bolometric light curve (top) and temperature measurements (bottom) from \citet{Tartaglia2016}.}
\end{figure}

The best-fit values of $R_e$, $M_e$, $v_e$ (or $v_s$) and $t_{\rm offset}$ from each model fit are presented in Table \ref{tab:results}. The best-fit models are presented in Figure \ref{fig:lc}, and the distribution of the parameters in Figure \ref{fig:corner}. The bolometric luminosity, radius and temperature evolution of the best-fit P15 and SW16 models are consistent with the measurements of \citet[][Fig. \ref{fig:lrt}]{Tartaglia2016}. 

\begin{figure*}
\includegraphics[width=\textwidth]{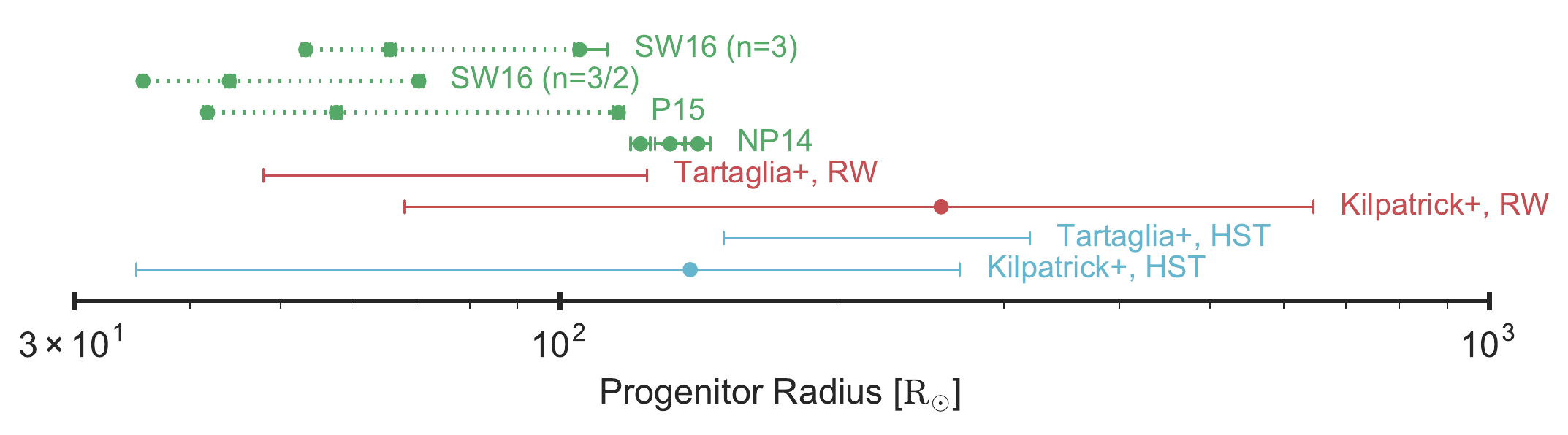}
\caption{\label{fig:radii}Radius estimates of the progenitor of {\gkg} from (bottom to top) {\it HST} imaging analysis of \citet{Kilpatrick2016b} and \citet[][considering both their progenitor candidates]{Tartaglia2016}, \citet{Rabinak2011} fits to the early light curve from \citet{Kilpatrick2016b} and to the early temperature evolution from \citet{Tartaglia2016}, and the three methods presented in this work for the low, nominal and high estimates of the host-galaxy extinction. The methods of NP14 and P15 rely on analytical approximations which are accurate up to factors of a few - a systematic uncertainty which is not plotted here.}
\end{figure*}

The P15 fit reproduces the decline from the first peak in all bands (dashed lines in Figure \ref{fig:lc}), but does not capture the sharp peak itself. One of the limitations of the P15 approach is that it does not consider the details of the stellar density profile, and thus might over-estimate the luminosity and under-estimate the temperature at the early rise to the first peak (P15). The SW16 fits, on the other hand, reproduce the shape of the first peak more accurately, but they miss some of the sharp decline in the UV bands.

The analytical approximations in NP14 and P15 claim to provide parameter estimates which are accurate to within a factor of a few. Within that accuracy, the results presented here are consistent with the results of \citet{Kilpatrick2016b} and \citet{Tartaglia2016} from both the {\it HST} pre-explosion imaging and their respective \citet{Rabinak2011} fits (Fig. \ref{fig:radii}). However, all models (for the three assumed host-extinction values) prefer smaller progenitor radii overall than those estimated by \citet{Tartaglia2016} from {\it HST} imaging.

The P15 fit converges on an expansion velocity $v_e$ which is within $\sim20\%$ of the value measured independently from the H$\alpha$ P-Cygni line in a spectrum taken during the shock-cooling peak \citep{Jha2016}. This is an important consistency check. 

However, the shock velocity $v_s$ preferred by the SW16 model fits is a factor of $\sim1.6$ larger than the H$\alpha$ expansion velocity measured by \citet{Jha2016} when in-fact $v_s$ is expected to be approximately half of the expansion velocity of some mass coordinate \citep{Matzner1999}. This discrepancy might be due to our use of the fiducial approximations to $f_{\rho}$ from SW16 (Eq. \ref{eq:sw16f}), which may not be accurate for the progenitor of {\gkg}. Similarly, the fiducial scaling factor we use for the temperature evaluation (Eq. \ref{eq:sw16temperature}) may differ from the true value appropriate for this event. 

Full hydrodynamical modeling of our data could allow for more accurate determinations of the properties of the progenitor of {\gkg}, and will be able to test the validity of the analytical approximations.

As wide-field transient surveys find more core collapse SNe at very early stages, and robotic telescopes obtain high-cadence multi-band followup during the first days since explosion, shock cooling models will allow for progenitor properties to be mapped statistically for SN populations. It is therefore important to be able to calibrate these models correctly in cases like {\gkg} where both early data and direct pre-explosion imaging exist.

~\\
We are grateful to N. Sapir for assistance interpreting the SW16 models. Support for IA was provided by NASA through the Einstein Fellowship Program, grant PF6-170148. DAH, CM, and GH are funded by NSF AST-1313484.. SJS acknowledges (FP7/2007-2013)/ERC grant 291222. DJS acknowledge NSF grant AST-1517649. This work makes use of observations from the LCOGT network. ATLAS observations were supported by NASA grant NN12AR55G. SOUSA is supported by NASA's Astrophysics Data Analysis Program through grant NNX13AF35G. This work made use of the NASA/IPAC Extragalactic Database (NED) which is operated by the Jet Propulsion Laboratory, California Institute of Technology, under contract with the National Aeronautics and Space Administration.


\end{document}